\documentclass[aip,
 sd,%
 amsmath,amssymb,
  preprint,%
]{revtex4-1}

\usepackage{graphicx}
\usepackage{dcolumn}
\usepackage{bm}

\usepackage{natbib}
\usepackage{units}
\usepackage{graphicx}
\usepackage{epstopdf}
\usepackage{hyperref}
\usepackage{color}
\usepackage[utf8]{inputenc}
\usepackage{empheq}
\usepackage{cases}
\usepackage{units}

\begin{document}

\preprint{AIP/123-QED}

\title[]{On the ions acceleration via collisionless magnetic reconnection in laboratory plasmas}

\author{E.~Cazzola}
\email{emanuele.cazzola@wis.kuleuven.be}
\affiliation{Center for mathematical Plasma Astrophysics, Department of Mathematics, K.U. Leuven (University of Leuven), Celestijnenlaan 200B, B-3001 Leuven, Belgium}

\author{D.~Curreli}
\email{dcurreli@illinois.edu}
\affiliation{Nuclear, Plasma and Radiological Engineering, University of Illinois at Urbana-Champaign, 216 Talbot Laboratory, 104 South Wright Street, Urbana, IL 61801, Illinois, USA}

\author{S.~Markidis}
\email{markidis@kth.se}
\affiliation{PDC Center for high Performance Computing, KTH Royal Institute of Technology, Teknikringen 14, 10044 Stockholm, Sweden.}

\author{G.~Lapenta}
\email{giovanni.lapenta@wis.kuleuven.be}
\affiliation{Center for mathematical Plasma Astrophysics, Department of Mathematics, K.U. Leuven (University of Leuven), Celestijnenlaan 200B, B-3001 Leuven, Belgium}

 \date{\today}
             
\begin{abstract}
 
This work presents an analysis of the ion outflow from magnetic reconnection throughout fully kinetic simulations 
with typical laboratory plasmas values.
A symmetric initial configuration for the density and magnetic field is considered across the current sheet.
After analyzing the behavior of a set of nine simulations with a reduced mass ratio and with a permuted value of three initial electron temperature and magnetic field intensity,
the best ion acceleration scenario is further studied with a realistic mass ratio in terms of the ion dynamics and energy budget.
Interestingly, a series of shock waves structures are observed in the outflow, resembling the shock 
discontinuities found in recent magnetohydrodynamic (MHD) simulations. 
An analysis of the ion outflow at several distances from the reconnection point is presented, in light of possible laboratory applications. The analysis suggests that magnetic reconnection could be used
as a tool for plasma acceleration, with applications ranging from electric propulsion to production of ion thermal beams.

\end{abstract}

\pacs{Valid PACS appear here}

\maketitle

\section{\label{sec:introduction} Introduction}

Magnetic reconnection is an important physical process occurring when two anti-parallel magnetic field lines encounter. Main feature of reconnection is the sudden restructuration of the magnetic field lines 
topology into a new more
stable and more favourable energetic configuration, accompanied with the release of the magnetic energy priorly stored in the reconnecting magnetic field lines. 
The released energy concurs
to heating and accelerating plasma particles, which makes magnetic reconnection a strongly kinetic process.
Historically, magnetic reconnection has been mainly studied in the astrophysical domains to explain explosive astrophysical events such as solar flares 
and geomagnetic substorms \citep[e.g.][]{birn2001,burch2009,kuznetsova1998,kuznetsova2001,gonzalez2016}.
In terrestrial applications, however, magnetic reconnection is encountered in fusion devices as responsible for magnetic island formation and tearing 
mode evolution, with detrimental consequences to plasma confinement and sustained fusion power production \cite{von1974,yu2014}. 
Nevertheless, simulations of the ion scale evolution with typical achievable plasma laboratory parameters is still particularly poorly represented in the
literature.
Previous studies on the reconnection ion dynamics have already been presented for the astrophysical cases \citep[.e.g][]{kuznetsova1996} 
and astrophysical reconnection explained with ad-hoc laboratory setup \citep[e.g.][]{yamada2016}, but to date poor knowledge is shared on the ion 
dynamics under suitable conditions for terrestrial purposes.
To further reduce this gap, in this work we present a parametric study from fully kinetic simulations of the ion acceleration from magnetic reconnection initially set with typical plasma laboratory values.
In particular, we aim at quantifying the effectiveness of magnetic reconnection for plasma acceleration and propulsion. With this purpose, 
the average outflow velocity across a specific cross section normal to the outflow direction is taken into consideration. From the latter, a specific quantity
is especially derived through the ratio of the average velocity and the gravitational acceleration constant $g_0$ (at sea level), in order to resemble the specific impulse 
traditionally used to distinguish among different plasma propulsion technologies.
After an initial set of nine runs for hydrogen plasma at a reduced mass ratio $m_r = 512$ with different magnetic field strenghts and electron temperatures, 
a more detailed analysis of the most promising case is 
performed by raising the mass ratio to the realistic value $m_r = 1836$.
Besides the energetic analysis of the reconnection outflow, we observe the formation of a set of flow shock discontinuities, which
partially remind those found by means of magnetofluidodynamics (MHD) simulations \cite{zenitani2011, zenitani2015}.

The paper is structured as follows: Section \ref{sec:simulations} gives an exhaustive description of the code used and all the configurations considered in this work. Interesting results are presented in 
Section \ref{sec:results}. Finally, Section \ref{sec:conclusions} will draw the most important conclusion, as well as point out possible future work directions.

\section{\label{sec:simulations} Simulation Setup}

All simulations have been performed using the fully kinetic massively parallel implicit moment particle-in-cell code iPIC3D \cite{markidis2010,innocenti2016}.
Simulations are addressed in 2.5 dimensions, meaning that all the vector quantities are three dimensional 
but their spatial variation is solved on a two dimensional plane. This choice allows for a better representation of the evolution projected 
on a specific plane
by maintaining the three dimensional nature of the problem.
The code makes use of the following normalizations: lengths are normalized to the ion skin depth $d_i = \nicefrac{c}{\omega_{p,i}}$, where $\omega_{p,i} = \left( \nicefrac{4 \pi n_i e^2}{m_i} \right)^{0.5}$ is
the ion plasma frequency, as well as the time unit in iPIC3D, velocities are normalized to the speed of light $c$, particles charge is normalized to the electric charge $e$ and 
mass is normalized to the ion mass $m_i$.

The initial configuration consists of a $ \left( 20 \times 12 \right)\ \unit{d_i}$ box with a single not-balanced current sheet.
In the simulations, a fixed cartesian frame of reference is adopted, with the $x$ coordinate being parallel to the current 
sheet, the $y$ coordinate being the direction of the magnetic field change, and the $z$ coordinate completing the reference set in the out-of-plane direction.
By considering a typical density value of 
$n = 10^{19}\ \unit{m^{-3}}$, it translates to a simulated box size of $\left( 144.2 \times 86.5 \right)\ \unit{cm}$ ($d_i = 7.2\ \unit{cm}$), which perfectly integrates with the traditional laboratory devices scale. 
While the initial density is kept constant all over the domain, together with the initial temperature, the magnetic profile across the current sheet follows the widely used hyperbolic function from the Harris 
equilibrium \cite{harris1962,birn2001}, which 
reads

\begin{equation} 
B_{x} \left( y \right) = B_0 \tanh{\left( \frac{y}{\lambda} \right)}
\end{equation}
where $\lambda$ is the current sheet half-thickness, here chosen as $0.5\ \unit{d_i}$, and $B_0$ is the asymptotic magnetic field intensity. 
This specific initial setup partially recalls the not-force-free simulation studied in \textcite{birn2009}. However, main difference between the two cases is the very low-$\beta$ 
attained with the values here considered (i.e. $\beta \sim 1.6 \cdot 10^{-4}$), compared to the high-$\beta$ configuration proposed in that analysis.
An initial perturbation is applied in the middle of the current sheet to trigger the reconnection process at a selected point.

Boundary conditions are chosen open along each boundary to better represent a realistic physical system. Particles and fields are free to escape the domain across the $x$ boundaries, 
together with a particles re-injection from the $y$ borders, similarly to what is done in \textcite{wan2008}. 
This approach allows for a significant reduction of the simulated domain,
although the choice of using open boundary conditions poses further remarkable differences in the reconnection outcome physics with respect to the traditional periodic 
boundary conditions \cite{daughton2007,wan2008}.

We performed a set of 9 simulations by maintaining fixed the ion temperature at $T_i = 0.025\ \unit{eV}$ (i.e. room temperature), and varying the magnetic field intensity and electron temperature 
according to the values summarized 
in table \ref{tab:runs}. 
Table \ref{tab:runs} also reports the corresponding values of the Alfv\'en and ion acoustic speeds, 
together with the plasma parameter $\beta$ as the ratio between thermal and magnetic energy.
The ion acoustic speed considers a polytropic index $1.67$ typical of ideal plasmas.
The initial particles velocity distribution is considered pure maxwellian with no initial drift.
The range of $B$, $T_e$ and $\frac{T_i}{T_e}$ values has been chosen within values easily achievable in laboratory using common magnets and heating systems. 
Also note the larger electron temperature with respect to the ion temperature, as normally encountered in the experimental practice.

To further save computational time, a reduced mass ratio $m_r = 512$ is considered. The space resolution adopted is $ dx = dy \sim 0.7\ \unit{d_e}$ ($d_e$ the electron skin depth)
whereas the temporal resolution is 
$dt \cdot \omega_{c,e} = 0.0369$ (runs $1$ through $3$), 
$dt \cdot \omega_{c,e} = 0.1182$ (runs $4$ through $6$) and
$dt \cdot \omega_{c,e} = 0.3692$ (runs $7$ through $9$),
depending on the specific run to limit the arise of numerical instabilities ($\omega_{c,e}$ the electron cyclo-frequency).
\begin{table}
\begin{centering}
  \begin{tabular}{| c | c | c | c | c | c | c |} 
  \hline

\textbf{Run} & $\mathbf{B\ \unit{\left[ Gauss \right]}}$ & $\mathbf{T_e\ \unit{\left[ eV \right]}}$ & $\mathbf{\frac{T_i}{T_e}}$ & $\mathbf{V_A\ \unit{\left[ \nicefrac{km}{s} \right]}}$ & $\mathbf{C_s\ \unit{\left[ \nicefrac{km}{s} \right]}}$ &  $\mathbf{\beta}$  \tabularnewline
\hline

 Run 1 & 200 & 0.5 & 0.05 & 137.9 & 8.95 & 0.005 \tabularnewline \hline
 Run 2 & 200 & 3 & 0.0083 & 137.9 & 21.91 & 0.0302 \tabularnewline \hline
 Run 3 & 200 & 10 & 0.0025 & 137.9 & 40.01 & 0.1 \tabularnewline \hline
 Run 4 & 800 & 0.5 & 0.05 & 551.5 & 8.95 & $3.15 \cdot 10^{-4}$ \tabularnewline \hline
 Run 5 & 800 & 3 & 0.0083 & 551.5 & 21.91 & 0.0019 \tabularnewline \hline
 Run 6 & 800 & 10 & 0.0025 & 551.5 & 40.01 & 0.0063 \tabularnewline \hline
 Run 7 & 5000 & 0.5 & 0.05 & 3446.9 & 8.95 & $8.06 \cdot 10^{-6}$ \tabularnewline \hline
 Run 8 & 5000 & 3 & 0.0083 & 3446.9 & 21.91 & $4.84 \cdot 10^{-5}$ \tabularnewline \hline
 Run 9 & 5000 & 10 & 0.0025 & 3446.9 & 40.01 & $1.61 \cdot 10^{-4}$ \tabularnewline \hline

\end{tabular}
\end{centering}
\caption{\label{tab:runs} Summary of the magnetic field $\mathbf{B}$ and electron temperature $T_e$ considered 
for the runs with reduced mass ratio $m_r = 512$. The Alfv\'en speed $V_A$, the ion acoustic speed $C_s$ and the plasma parameter $\beta$ are also provided.}
\end{table}
The last simulation consists of the same configuration as Run 9 in table \ref{tab:runs} (i.e. $B = 5000\ \unit{Gauss}$ and $T_e = 10\ \unit{eV}$) with a realistic mass ratio 
$m_r = 1836$. With respect to Run 9, the temporal step is now $dt \cdot \omega_{c,e} = 0.662$ 
and the space resolution is $dx = dy \sim 1.33\ \unit{d_e}$.

\section{\label{sec:results} Results}

\begin{figure}
 \begin{center}
  \includegraphics[scale=0.4]{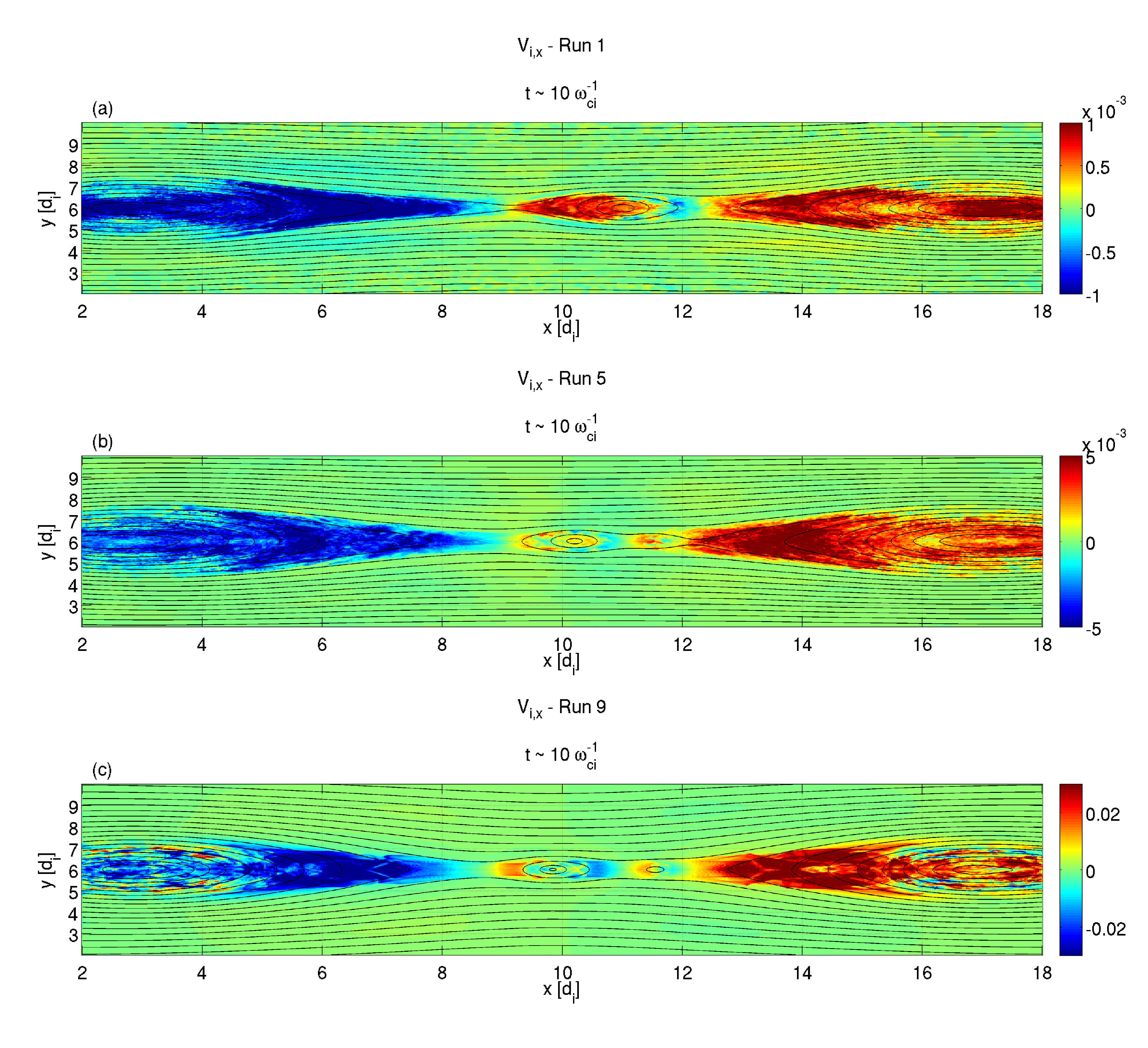}
    \caption{Plots of the $x$-component of the ion velocity, normalized to the light-speed, for three representative runs from table \ref{tab:runs},
    namely Run 1, Run 5 and Run 9, at the same time step. Black lines in background represent the magnetic field lines.}\label{fig:VxionsRuns}
 \end{center}
\end{figure}

This section reports the results from the simulations of section \ref{sec:simulations}. 
Figure \ref{fig:VxionsRuns} shows the $x$-component of the ion outflow (drift velocity)
for three representative runs (i.e. Run 1, Run 5 and Run 9) from table \ref{tab:runs} 
at the same time frame ($t = 10\ \unit{\omega_{c,i}^{-1}}$), to highlight
the different evolution of magnetic reconnection under different initial conditions.
An increasing outflow velocity is observed as both the temperature and magnetic field increase. 
In particular, the highest outflow velocity is observed for Run 9 (Fig \ref{fig:VxionsRuns}.c), in condition of high magnetic field and high
electron temperature. 
Also, the reconnection evolution is observed to be particularly rapid, being fully developed already at $10\ \unit{\omega_{c,i}^{-1}}$, as can be seen
from Run 9 where reconnection is at later stages of development compared to cases of weaker $B$-field (e.g. the Run 1 and 3). 

The formation of secondary islands is observed all over the current sheet, as a conseguence of the initial chosen not-equilibrium (some
reminiscence is visible at the edges of the simulated box e.g. at $x \sim 17\ \unit{d_i}$ in figure \ref{fig:VxionsRuns}).
These islands show  to have a small amplitude and are being completely wiped out by the main reconnection outflow from the 
initially perturbated region.  The latter underlines the importance of applying an initial localized perturbation to trigger the process at a preferred location. 
With no initial perturbation, 
the process would start at few randomly distributed reconnection locations along the current sheet, 
growing and evolving irregularly until ultimately interacting 
and merging together to form larger magnetic islands, as explained e.g. in \textcite{drake2006}, \textcite{pritchett2008b}, \textcite{oka2010} and \textcite{cazzola2015, cazzola2016}.

Additionally, larger magnetic islands are seen to form and stand
near to the region where the reconnection takes origin. This effect is commonly found since the earliest simulations (e.g.
\textcite{daughton2006}) and is
not seen to be influencing the following analysis, 
provided that only cross-sections far from this region are considered.

\subsection{Comparison with Simplified Models of the Ion Outflows}

\begin{figure}
 \begin{center}
  \includegraphics[scale=0.4]{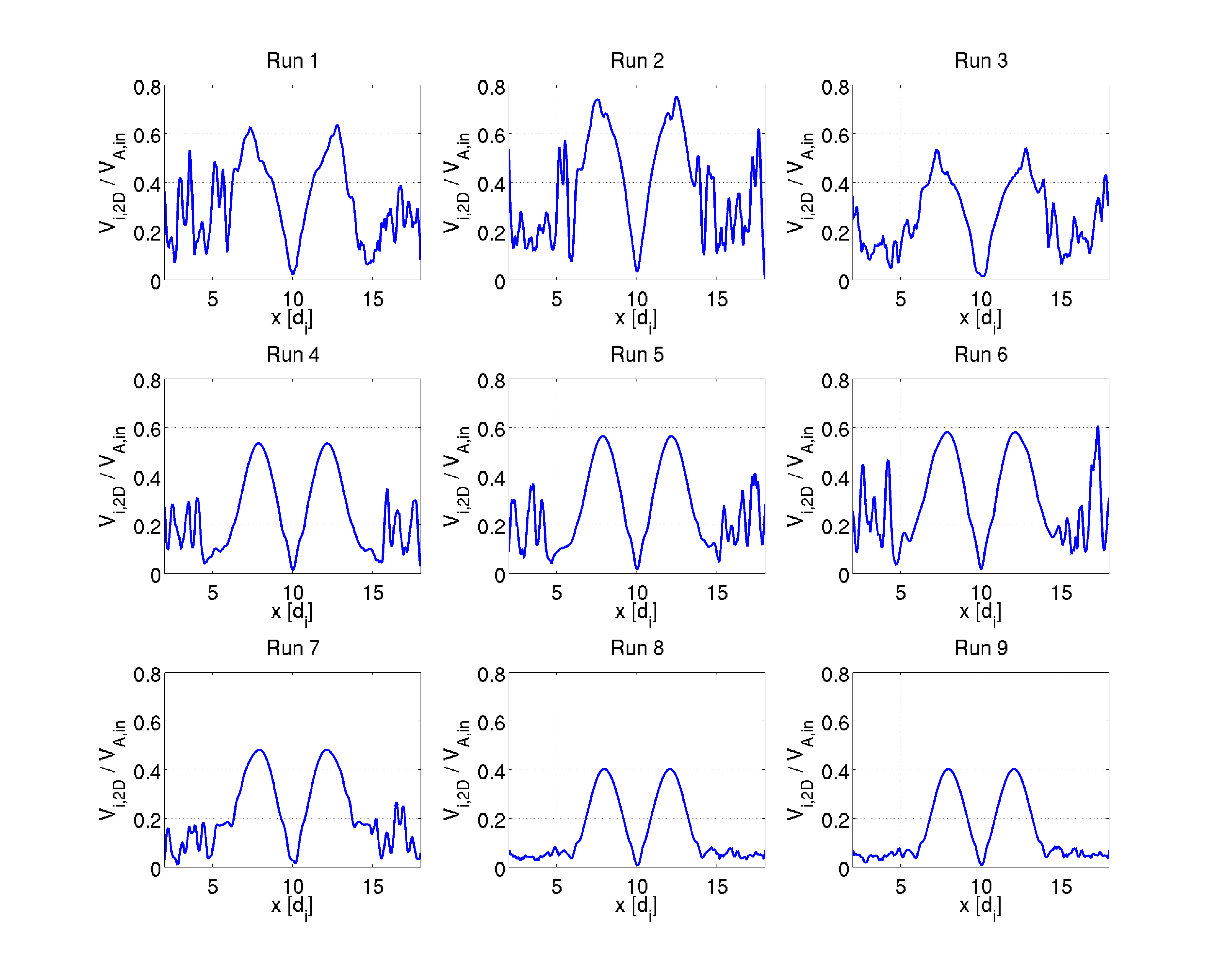}
    \caption{In-plane ion velocity profiles along the current sheet for the runs in table \ref{tab:runs}. Values are computed from simulations by only considering the $x$ and $y$ component, and are normalized to the 
    inflow Alfv\'en speed.}\label{fig:VxiVa}
 \end{center}
\end{figure}

This section aims at comparing the ion velocity obtained from the simulations with that from existing theoretical models, such as those described in \textcite{priest2007}
and \textcite{simakov2008}.
Figure \ref{fig:VxiVa} shows for each run the ion outflow profiles along the current sheet normalized to the inflow Alfv\'en speed, as computed considering only the in-plane components 
(i.e. $V_x$ and $V_y$)
to address a better comparison with pure 2D models.
Despite this approximation,
we noticed an insensibile influence of the $z$-component compared to the $2D3V$ case.
The inflow Alfv\'en speed has been evaluated at the location where the three characteristic velocities
$V_{i,y}$, $V_{e,y}$ and $\frac{\left( \mathbf{E} \times \mathbf{B} \right)}{B^2} \cdot \mathbf{\hat{e}_y}$ begin to diverge (i.e. $6.8 \le y \le 9.1\ \unit{d_i}$, depending on the Run and
time considered), whereas
the time window was chosen to follow the reconnection region before the formation of the central island.
The analysis reveals a maximum velocity outflow nearly half as fast as the inflow Alfv\'en speed, and equally developed over all runs, followed by a steep velocity drop  at the
edge caused by the formation of a dipolarization
front \cite{lapenta2014b,sitnov2009,sitnov2011}. Results are charted in table \ref{tab:predVel}, second column. 

Even though considered at its very first stages, magnetic reconnection develops
a series of loss channels causing a remarkable velocity drop from the theoretical Alfv\'enic value.
However, still noticeable is the nearly constant outflow value held over all the cases, with values up to  $0.75\ \unit{V_A}$ (i.e. Run $2$).

In the same table we compare these results with those from the model proposed by Sweet and Parker \cite{priest2007} (fourth and fifth columns) 
and the recent model in \textcite{simakov2008} (third column). 
The basic Sweet-Parker model states that the output velocity is Alfvenic when
the $\nabla \cdot \mathbf{P}$ term is neglected in the equation of motion. On the other hand, the reconnection outflow can be either slowed down further 
or accelerated by the presence of a pressure gradient 
along its ouflow direction. 
An updated model was then proposed including the pressure gradient \citep{priest2007}, and used here for comparison. The results for our configuration
are shown in column four.
In particular, the pressure difference between the outflow and the inflow region has been considered for the assessment. 
Likewise, plasma compressibility can also play an important role at influencing the outflow velocity. 
Its effect has been here taken into account in a separate evaluation given in the fifth column.

\begin{table}
\centering
\resizebox{\textwidth}{!}{\begin{tabular}{| c | c | c | c | c | c |}
 \hline
\textbf{Run} & Velocity from simulations & \textcite{simakov2008} model & Sweet-Parker model with pressure gradient & Sweet-Parker model with compressibility  \tabularnewline
\hline

 Run 1 &  $0.65$ & $0.550$ & $ 1.4023 $ & $ 0.607 $  \tabularnewline \hline
 Run 2 &  $0.74$ & $0.230$ & $ 1.4006 $  & $ 0.267 $  \tabularnewline \hline
 Run 3 &  $0.55$ & $0.314$ & $ 1.4005 $ & $ 0.4572 $ \tabularnewline \hline
 Run 4 &  $0.52$ & $0.154$ & $ 1.4075 $ & $ 0.7951 $ \tabularnewline \hline
 Run 5 &  $0.56$ & $0.200$ & $ 1.4081 $ & $ 0.8085 $ \tabularnewline \hline
 Run 6 &  $0.58$ & $0.112$ & $ 1.4088 $ & $ 0.3236 $ \tabularnewline \hline
 Run 7 &  $0.49$ & $0.031$ & $ 1.4050 $ & $ 0.4075 $ \tabularnewline \hline
 Run 8 &  $0.40$ & $0.063$ & $ 1.4118 $ & $ 0.4366 $ \tabularnewline \hline
 Run 9 &  $0.40$ & $0.060$ & $ 1.4115 $ & $ 0.4364 $ \tabularnewline \hline

\end{tabular}}
\caption{\label{tab:predVel} Comparison of the ion velocity magnitude obtained from simulations (the out-of-plane component 
is not considered), with the model proposed in \textcite{simakov2008} (third column) and those proposed by Sweet and Parker for the case with, 
respectively, pressure gradient (fourth column) and plasma compressibility 
(fifth column). All values are normalized to the inflow Alfv\'en speed evaluated as explained in the manuscript.}
\end{table}

The analysis shows a qualitative agreement with the Compressible Sweet-Parker model and, to a lesser extent, with the
\textcite{simakov2008}'s model, with the latter mostly limited to the first three cases with the lowest field intensity.
The opposite is observed for the Sweet-Parker model considering the pressure gradient. 
Even though the outflow pressure results remarkably greater than the inflow pressure (not shown here for conciseness), such gradient 
is not significantly affecting the outflow velocity. On the contrary, the model predicts a super-alfv\'enic outflow velocity 
in line with the $\sqrt{2} V_A$ limit 
explained in \textcite{priest2007} for low $\nabla \cdot \mathbf{P}$. 
The relevant difference between this and the other models can be explained by the initial setup adopted:
despite the large pressure difference between inflow and outflow, 
the thermal energy ($p_P \propto n T \sim 10^{-6}\ \unit{\nicefrac{J}{m^-3}}$) 
is initially much lower than the magnetic field energy ($p_M \propto B^2 \sim 10^{-3}\ \unit{\nicefrac{J}{m^-3}}$). Being the Alfv\'en speed dependent 
on $\mathbf{B}$, and the pressure term dependent on the temperature, the remarkable field intensity and low temperature adopted here makes 
the pressure term much less dominant than the Alfv\'en speed term in the balance. 
Vice versa, the Sweet-Parker model including the  compressibility instead results  in qualitative 
agreement with the numerical simulations, with a satisfactory accuracy ($< 10 \%$) for 
Runs 7 to 9, where the magnetic field becomes more intense.
Finally, the discrepancy between the Simakov - Chac\'on model \cite{simakov2008} and the simulations can be explained by the intrinsic assumptions 
made in the model, which cause it to
fail when the magnetic pressure far exceed the plasma pressure, as occurring in low-$\beta$ laboratory plasmas.

\subsection{Results with a Realistic Mass Ratio}

Figure \ref{fig:VxionsRuns} shows that the highest outflow velocity is reached in Run $9$ (at higher B-field and $T_e$). 
This  Run $9$ is therefore taken as reference
for a further analysis with a realistic mass ratio $m_r = 1836$.
Notice that Table \ref{tab:predVel} instead states differently: this difference
is due to the inflow Alfv\'en speed to which the velocity values are normalized, which depends on the different inflow conditions.
The temporal step is now set at
$dt \cdot \omega_{c,e} = 0.662$
and we extended the global simulation time to observe the process over its whole evolution.

\begin{figure}
 \begin{center}
  \includegraphics[scale=0.50]{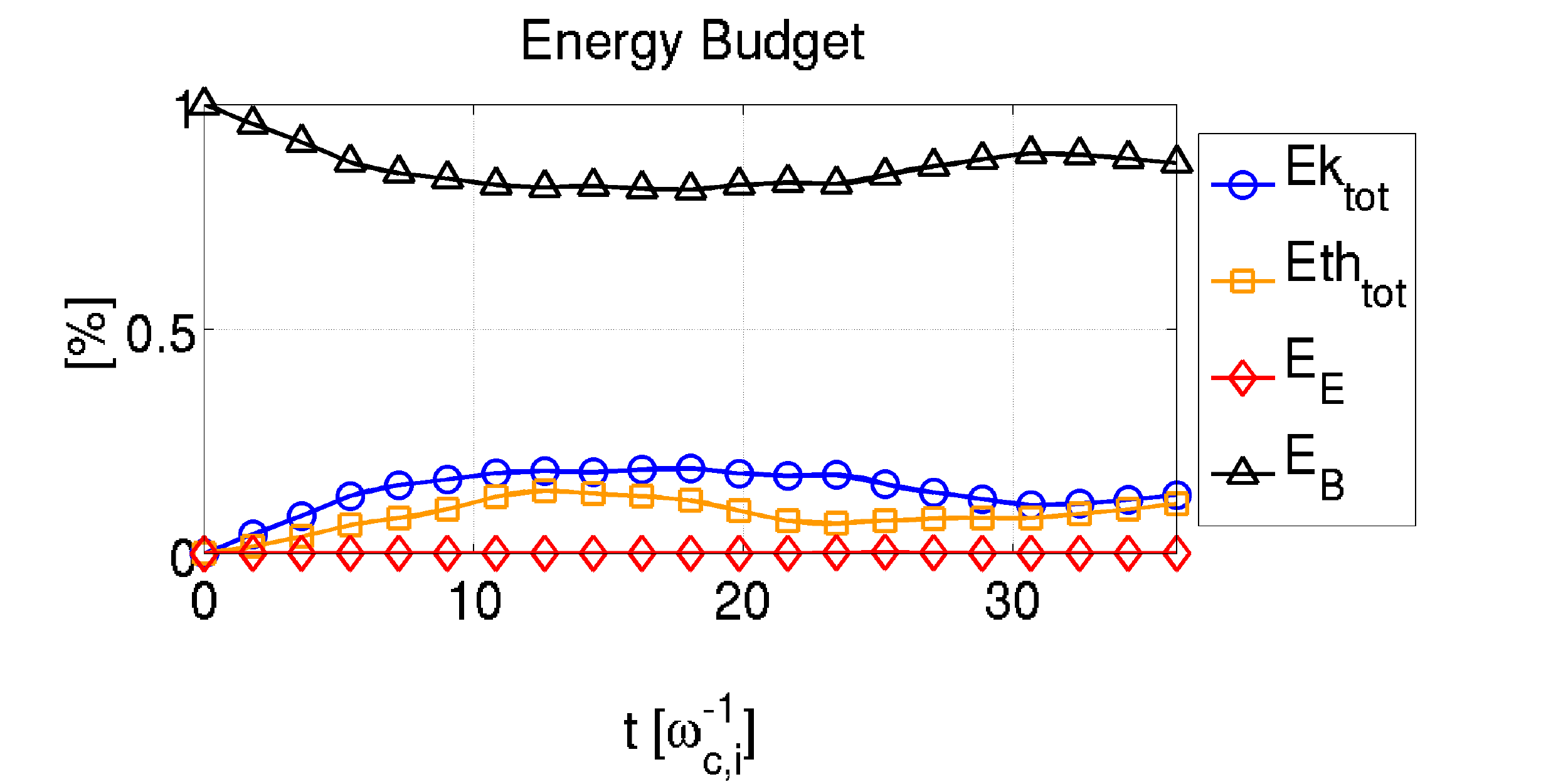}
    \caption{Temporal profiles of the EM fields energy ($E_B \text{ and } E_E$) and particles energy ($E_{K,tot}$) after magnetic reconnection. 
    $E_{K,tot}$ represents the total energy acquired by particles (i.e. $E_{K,tot} = E_d + E_{th}$, where 
    $E_{th}$ is the thermal energy component. All the values are normalized to the total energy $E_{tot} = E_B + E_E + E_{K,tot}$ at each time.} \label{fig:CQ}
 \end{center}
\end{figure}

First of all, in Figure \ref{fig:CQ} we show the temporal energy budget shared after the reconnection event between the electromagnetic fields (EM) and particles. $E_B \text{ and } E_E$ 
represent the magnetic and electric energy respectively, whereas $E_{K,tot}$ represents the total energy acquired by particles, such that $E_{K,tot} = E_d + E_{th}$, where 
$E_{th}$ is the thermal energy and $E_d$ the drift energy (not shown in Fig. \ref{fig:CQ}). These values are normalized to the totale energy $E_{tot} = E_E + E_B + E_{K,tot}$ at each time step. 
The magnitude at $t = 0$ yields $E_B >> E_{K,tot} \text{, } E_E$.
Besides neglecting the electric field energy, we observe that up to $25 \%$ of the initial magnetic energy is converted to particles energy during the first reconnection stages, with the thermal energy increasingly dominating 
the budget, to indicate that plasma is initially accelerated and later heated. A steady budget is then reached up to $\sim 25\ \unit{\omega_{c,i}^{-1}}$. 
The magnetic energy increase reported after $\sim 25\ \unit{\omega_{c,i}^{-1}}$ is believed to be caused by the 
formation of a relevant magnetic island.

\begin{figure}
 \begin{center}
  \includegraphics[scale=0.60]{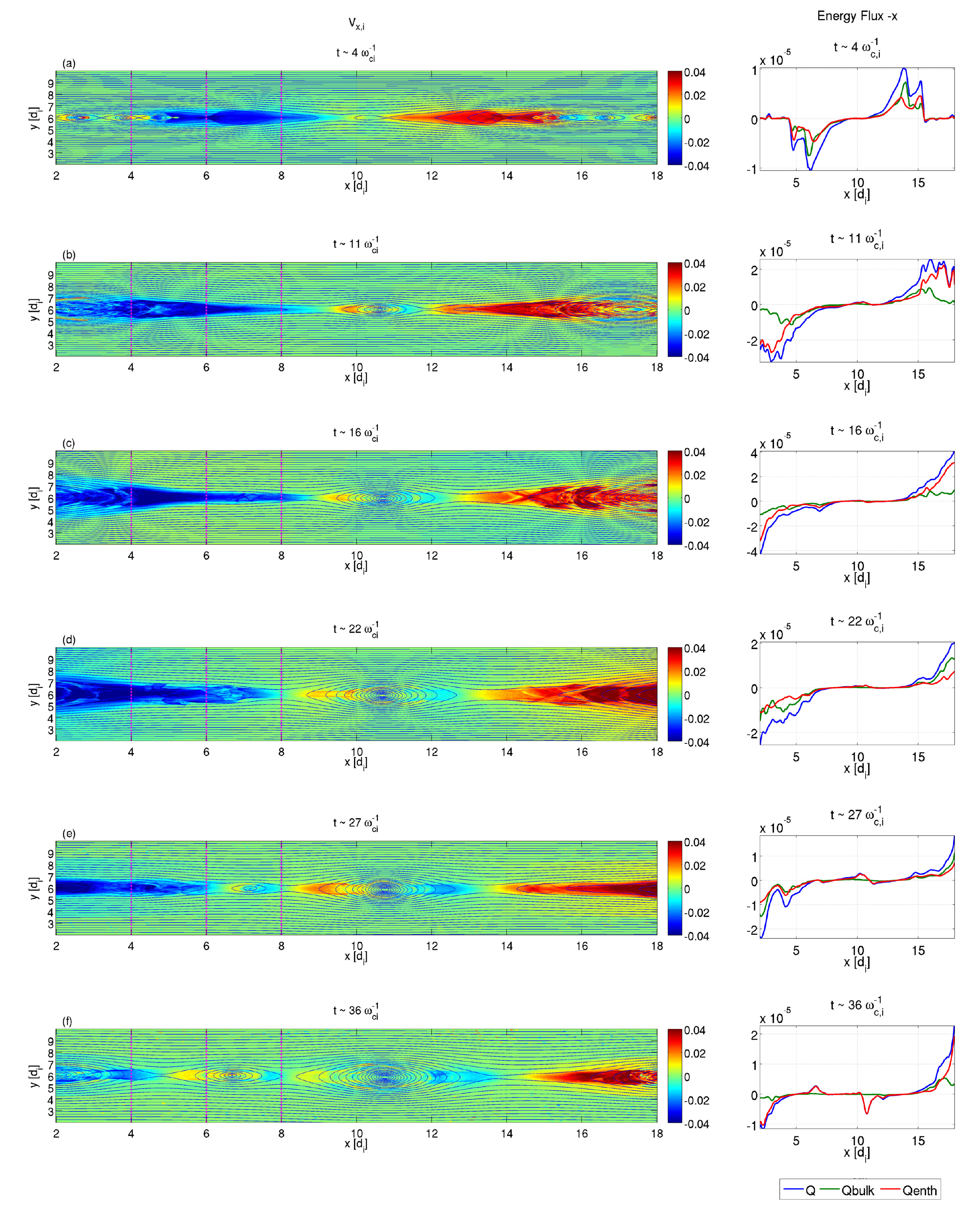}
    \caption{Plots of the $x$-component of the ion velocity at different times for the simulation with realistic mass ratio and configuration as Run 9 in table \ref{tab:runs},
    together with the energy analysis profiles as from equations \ref{eq:Qb}-\ref{eq:Q}. Velocity is normalized to the light-speed, while energy quantities 
    are qualitatively shown in code units. The purple lines represent the cut considered for the ion outflow analysis.}\label{fig:VxionsRuns1836}
 \end{center}
\end{figure}

Figure \ref{fig:VxionsRuns1836} shows instead the $x$ component of the ion velocity for this new configuration, together with the $y$-integrated profiles for the $x$-component of 
the bulk energy flux $\mathbf{Q_b}$, enthalpy energy flux $\mathbf{Q_{enth}}$
and their sum $\mathbf{Q}$ in the Eulerian frame, which read

\begin{equation} \label{eq:Qb}
 \mathbf{Q}_b  =  \mathbf{u} U_b \text{,   } U_b = \frac{mnu^2}{2} 
 \end{equation}
 
  \begin{equation} \label{eq:Qth}
 \mathbf{Q}_{enth}  =  \mathbf{u} U_{th} + \mathbf{u} \cdot \mathbb{P} \text{,   }  U_{th} = \frac{Tr \mathbb{P}}{2} 
\end{equation}

  \begin{equation} \label{eq:Q}
\mathbf{Q} = \mathbf{Q}_b + \mathbf{Q}_{enth}
\end{equation}
where $U_b$ the bulk kinetic energy, $m$ the mass and $n$ the density, $\mathbf{u}$ the directional bulk velocity vector,
$U_{th}$ the thermal energy and $\mathbb{P}$ is the pressure tensor.

The profiles show an overall dominance of the enthalpy flux over the bulk energy flux, whose outcome highlights 
how the initial energy released by the reconnection event and quasi-instantly transferred to the plasma is soon being turned into enthalpy. 
This result was also pointed out in \textcite{birn2009}, where 
both PIC and MHD simulations confirmed that in low-$\beta$ reconnection the enthalpy flux tends to dominate over the bulk energy flux, 
mainly generated by fluid compression.

\begin{figure}
 \begin{center}
  \includegraphics[scale=1.]{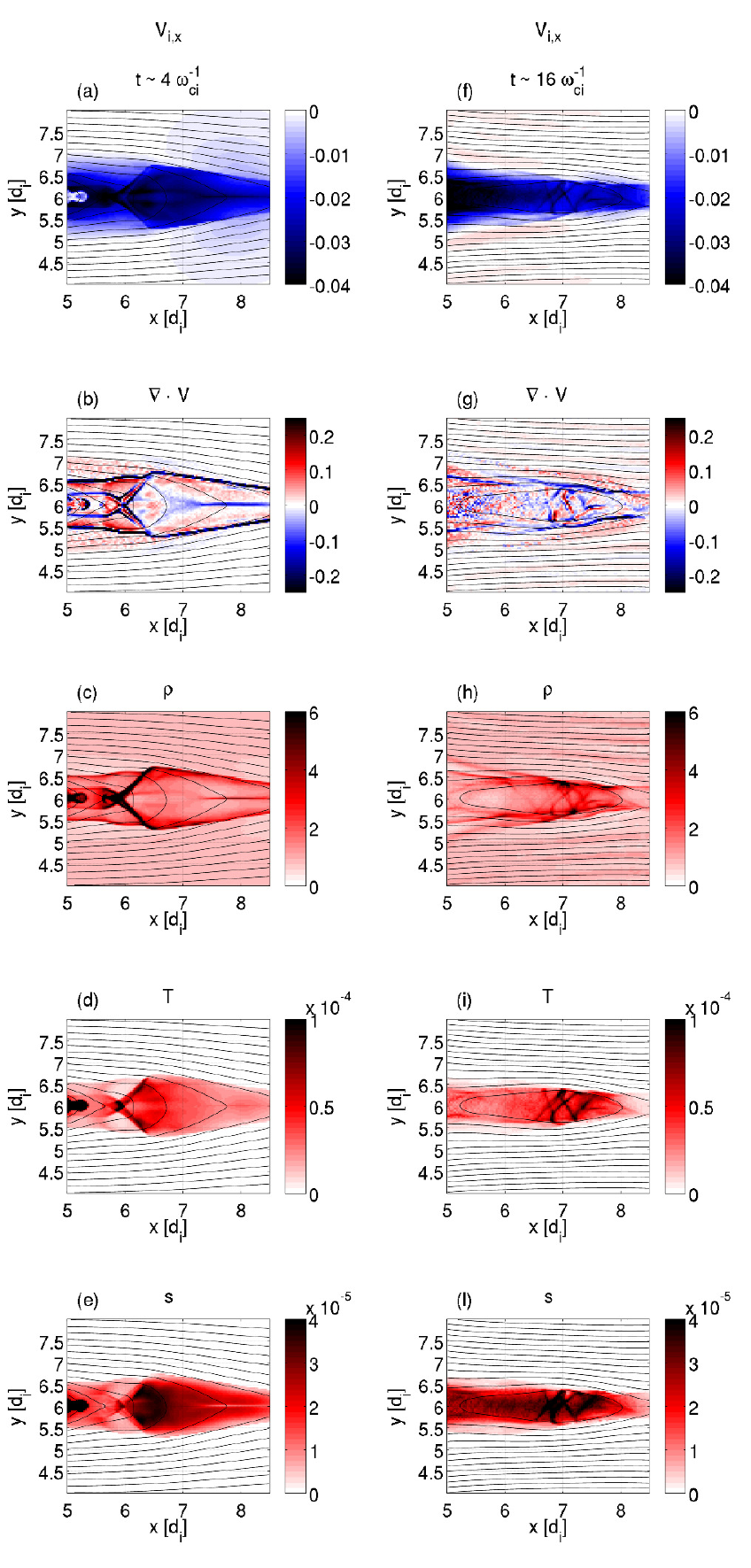}
    \caption{Situation at the reconnection outflow in Fig. \ref{fig:VxionsRuns1836} panel (a) represented by different fluidodynamic quantities: 
    ion velocity along $x$, convective derivative of the total velocity $\nabla \cdot V$, total mass density $\rho$, temperature $T$ and entropy $s$ (colorbar in code units).}\label{fig:outflowAnalysis}
 \end{center}
\end{figure}

In Figure \ref{fig:VxionsRuns1836} additional interesting features are observed in the reconnection outflows,
such as those discontinuity structures, e.g.  at $4\ \text{and } 16\ \unit{\omega_{c,i}^{-1}}$, which
recall those found and analyzed in \textcite{zenitani2011} and \textcite{,zenitani2015} for low $\beta$ plasmas.
A magnification of the leftside outflow at these time steps is then given in 
figure \ref{fig:outflowAnalysis}, where some important quantities are displayed, namely the convective derivative
$\nabla \cdot \mathbf{V}$, density $\rho$, temperature $T$ and the entrophy $s$. The latter quantity has been computed from the polytropic relation 
$s = \frac{p}{\rho^{\gamma}}$, where $p$ pressure. By assuming the plasma an ideal gas, the polytropic index is chosen as $\gamma = 1.67$.

At earlier times (panels (a)-(e)),
the $\nabla \cdot \mathbf{V}$ quantity reveals complex symmetric and regular discontinuity structures, which are only partially replicated in the temperature and entropy 
plots. 
In particular, the separatrices appear to be
remarkably highlighted due to the repentine change of the particles flow across them, as pointed out in \textcite{lapenta2014, Lapentabook}. The density also shows an increased value, 
unlike temperature and entropy which are not seen particularly marked. 
This situation resembles the Petschek-like slow shocks pointed out in \textcite{zenitani2011}, and recently found in kinetic simulations \cite{innocenti2015}. 
Interestingly is also the horizontal feature seen at $y = 6\ \unit{d_i}$ straightly 
coming out from the reconnection region, and visible  over all the plots. Unlike what seen in \textcite{zenitani2011}, 
whereby this signature coincides with a marked density cavity, in our case
the density in this region appears to be larger, together with an increase in the temperature and a decrease in entropy with respect to the surrounding plasma.
Noticeable are also the diagonal structures observed between $x = 5 \text{ and } 7\ \unit{d_i}$, which show a remarkable temperature and density growth. 
Such structures are also highlighted by $\nabla \cdot \mathbf{V}$, revealing the presence of a very complex symmetric structure upstream.
Furthermore, right behind the oblique discontinuities the magnetic field is noticed to repentinely change its direction by turning almost vertical, which may be 
a clear signature of a rotational discontinuity accompanied with a shock.
From the larger view in figure \ref{fig:VxionsRuns1836},
we notice the outer region of the structure to be indeed describing the interaction between the outflow and a growing magnetic island standing near the reconnection point.
Finally, a strong entropy rise is observed downstream in between the 
two oblique discontinuities.

A more irregular situation is instead observed later in time at $16\ \unit{\omega_{c,i}^{-1}}$. Particularly noticeable is the 
series of oblique discontinuities observed within the closed magnetic field. This structure clearly resembles the diamond-chain described in 
\textcite{zenitani2011} and \textcite{zenitani2015}. The structure is seen to fade out while flowing out, disappearing before reaching the island center. 
This structure is especially highlighted 
by an increase in both temperature and entropy, while the density appears to only increment weakly, with the maximum values observed on the magnetic island borders. 
Finally, the separatrices are again well remarked by $\nabla \cdot \mathbf{V}$, as expected, 
whereas the straight horizontal line seen at the earlier time is now no longer reported.

\begin{figure}
 \begin{center}
  \includegraphics[scale=.6]{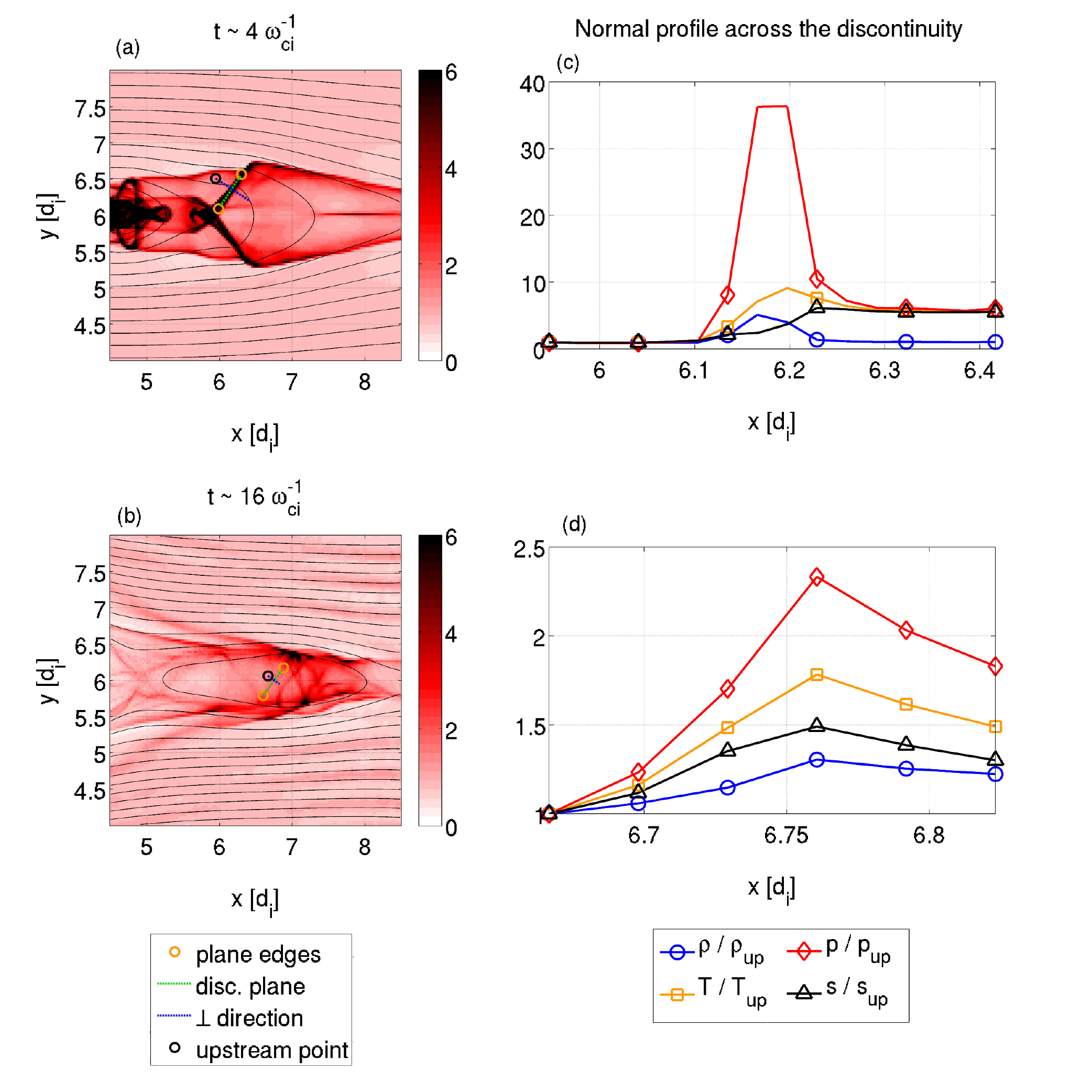}
    \caption{Plasma thermodynamic quantities ($\rho$ - mass density, $p$ - pressure, $T$ - temperature and $s$ - entropy) across two selected shock 
    discontinuities observed in 
    Fig. \ref{fig:outflowAnalysis}. All quantities are normalized to the corresponding value upstream, marked with a black circle. Figures (a), (c) 
    exhibit a behavior typical of rarefaction shock waves; figures (b), (d) of compressive shock waves.}\label{fig:shocks}
 \end{center}
\end{figure}

Even though a more in-depth analysis would be required (e.g. throughout a Wal\'en test), 
a first insight into the nature of these discontinuities can already be inferred by analysing the evolution of some relevant quantities across it.
Figure \ref{fig:shocks} shows the evolution of temperature, pressure, entropy and density along the direction perpendicular to the 
discontinuity plane for the two time frames reported in 
figure \ref{fig:outflowAnalysis}. The profiles are normalized to the upstream value marked
with a black circle in panels 6.a and 6.b. Notice that in the abscissa of panels 6.c and 6.d 
the projection on the $x$ coordinate has been preferred to the 
normal arc-length for more clarity. 
Two particular discontinuities are considered for the study, namely the sharp V-shape seen at $t \cdot \omega_{c,i} \sim 4$ and one of 
the visible jumps constituting the diamond-chain observed 
at $t \cdot \omega_{c,i} \sim 16$. In the first case, we observe a pressure increase, as well as an increase in the entropy and temperature. 
In particular, the density ratio slightly decreases across the discontinuity, with the temperature ratio
being larger than the density ratio. 
Such evolution exhibits the typical behavior of a rarefaction wave \cite{baumjohann1996}.  
The situation reverses at time $t \cdot \omega_{c,i} \sim 16$, where a compressive shock wave is encountered. At these later times, all the quantities increase
across the discontinuity. The temperature ratio is lower than the density ratio, and the downstream density is greater than upstream.
These features highlight a behavior typical of compressive shocks \cite{baumjohann1996}.

\begin{figure}
 \begin{center}
  \includegraphics[scale=0.35]{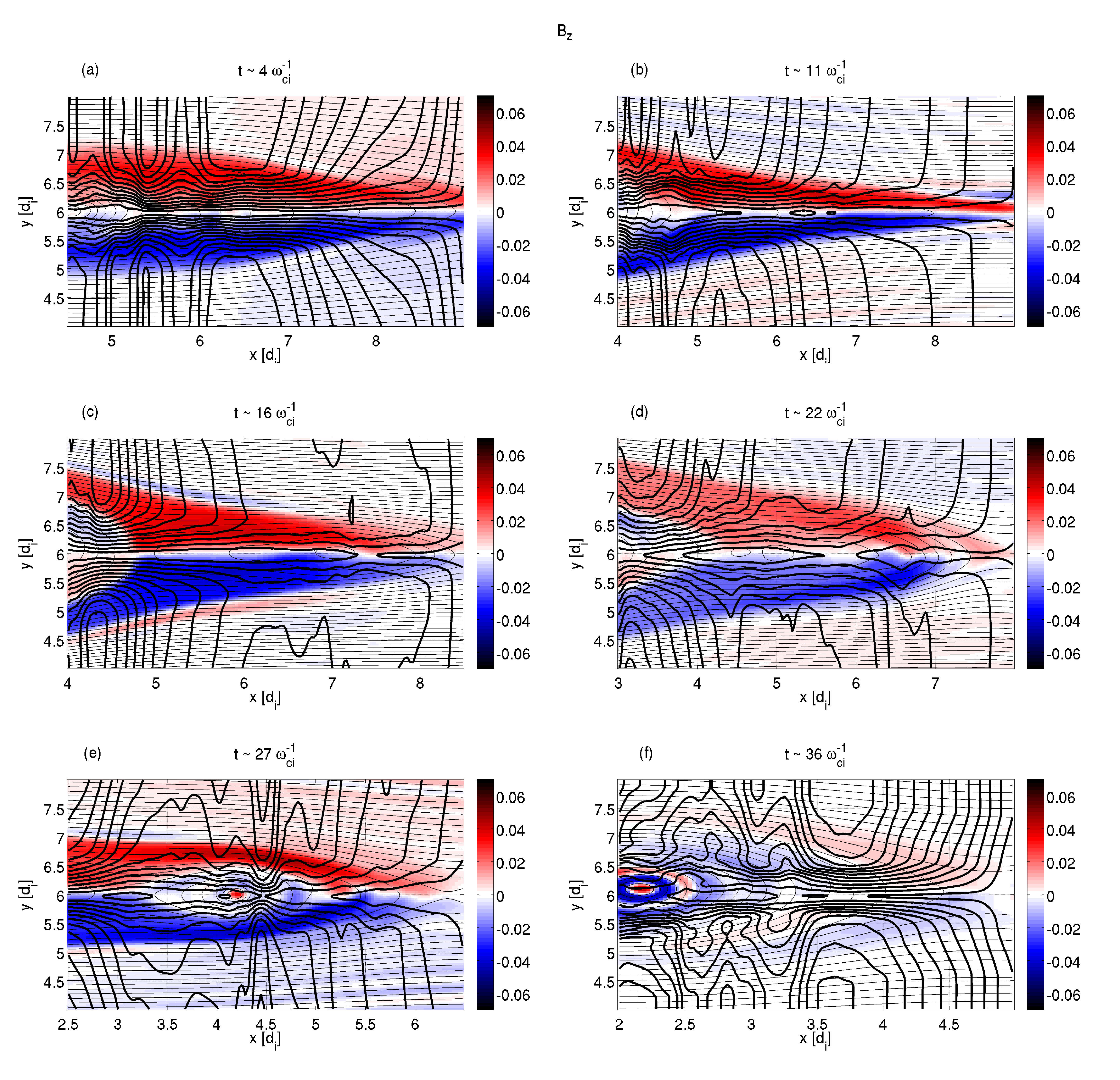}
    \caption{Zoom of the left reconnection out-of-plane 
    magnetic field $B_z$ (in code units) at different times for the simulation with realistic mass ratio, 
    overlapped with the Lagrangian ions trajectory (black thick lines) and
    magnetic field lines (black thin lines in background).}\label{fig:LagrVel}
 \end{center}
\end{figure}

To better understand the ions dynamics, Figure \ref{fig:LagrVel} shows the ions lagrangian trajectory over the lefthand outflow region at different temporal steps, 
overplotted upon the out-of-plane magnetic field $B_z$ and the magnetic field isocontour (black thin lines). 
The $B_z$ colorscale is kept contant over the plots, while the subdomain is shifted along $x$ according to the outflow motion to leave out the central growing island.
This type of representation for the ions dynamics has already been established in the literature as a good approximation for 
the trajectory undergone by a single fluid volume \cite{lapenta2014}.
From the plots we notice that, at the first simulation stages, the ions approaching the farthest outflow regions are prone to being bounced out 
soon after crossing the separatrices.
This effect has been observed to often occur in the presence of a magnetic island. 
However, most of the ions entering the domain from the vicinity of the reconnection region still tend to cross the separatrices uninfluenced and be then deflected outwards along the outflow,
in good agreement with other analyses reported in literature \citep{lapenta2014}.

\subsection{Ions Momentum}

The ions dynamics is further studied through the analysis of the average outflow velocity observed across the series of cross sections marked with a purple vertical lines in figure \ref{fig:VxionsRuns1836}.
In analogy with space applications, the average velocity is divided by the 
gravitational constant to retrieve what in plasma propulsion is known as specific impulse, which is simply

\begin{equation} 
I_{sp} = \frac{\overline{V}_x}{g} 
\end{equation}
where $g = 9.81\ \unit{m s^{-2}}$ is the gravitational acceleration at ground level and $\overline{V_x}$ is the average value of the $x$ velocity component.
The $I_{sp}$ unit is then seconds.

\begin{figure}
 \begin{center}
  \includegraphics[scale=0.60]{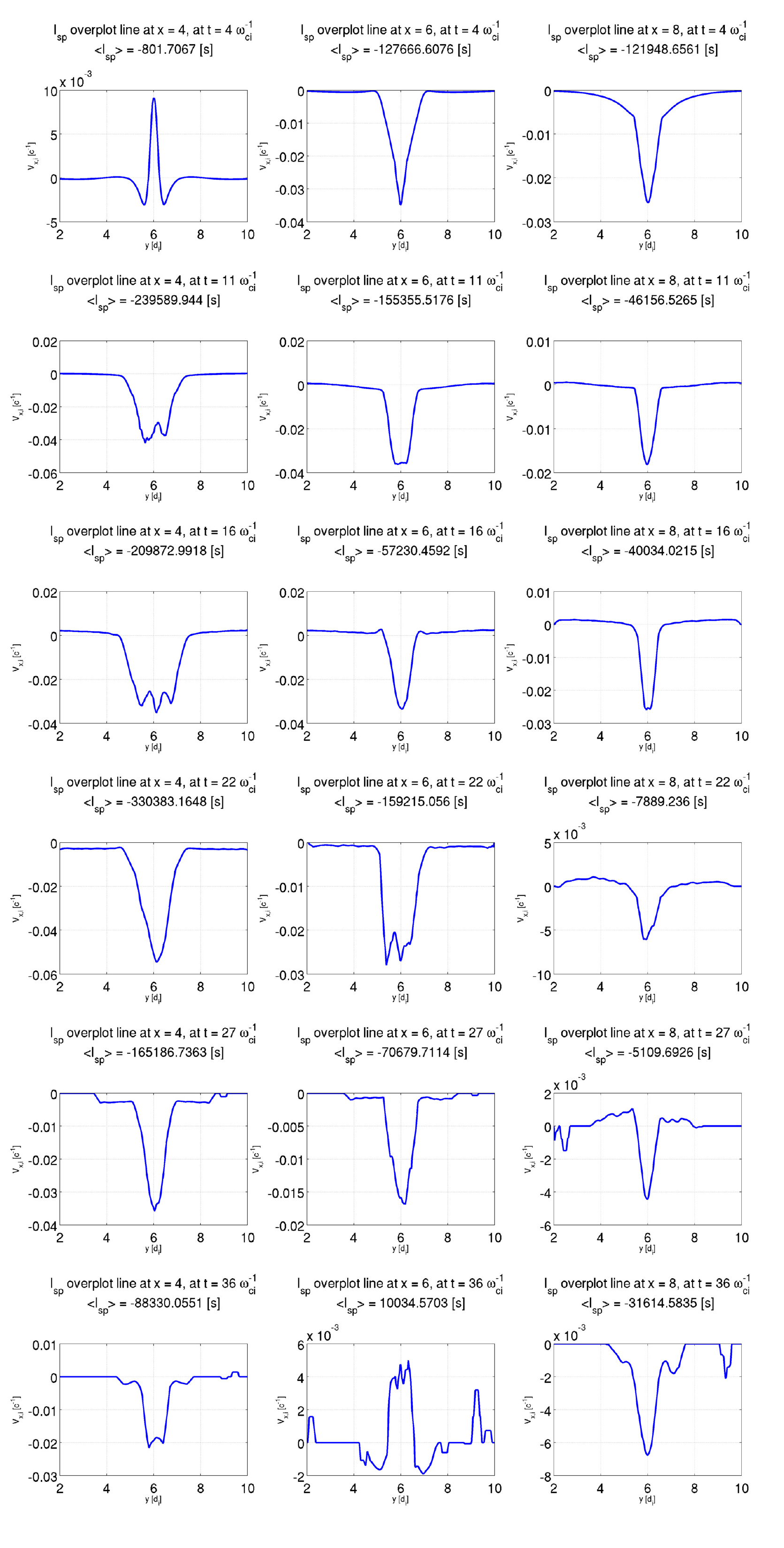}
    \caption{Profiles of the ion velocity along $x$ over the cut indicated with purple lines in Fig. \ref{fig:VxionsRuns1836}. 
    The mean value of the quantity $I_{sp}$ is shown at the top of each panel.}\label{fig:Isp1836}
 \end{center}
\end{figure}

\begin{figure}
 \begin{center}
  \includegraphics[scale=0.5]{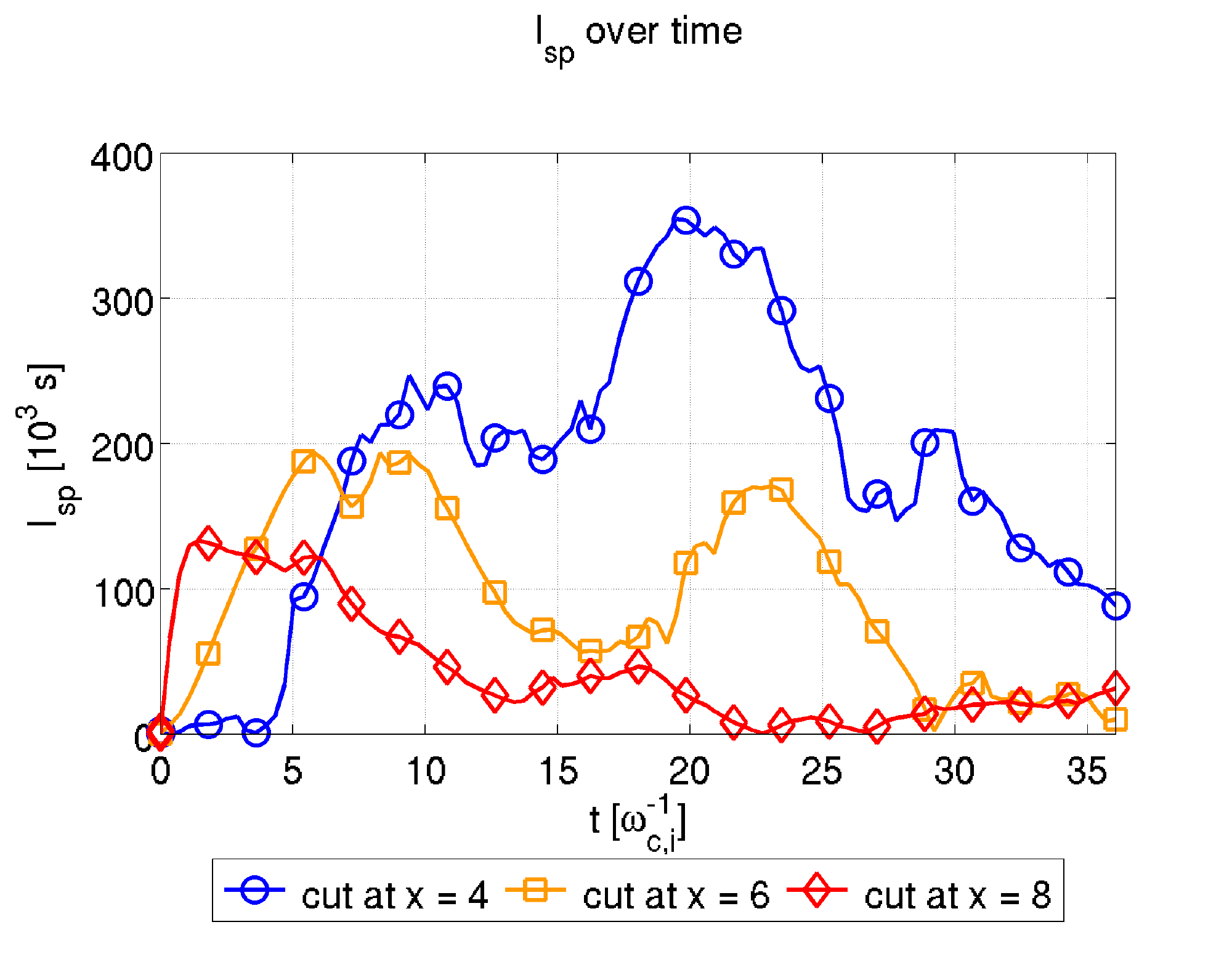}
    \caption{Temporal evolution of the quantity $I_{sp}$ over the whole simulation at the different cuts shown in Fig. \ref{fig:VxionsRuns1836}.}\label{fig:IspTime_mr1836}
 \end{center}
\end{figure}

Three different cross sections are considered over the lefthand outflow shown in figure \ref{fig:VxionsRuns1836} (dashed purple lines in figure). 
The resulting velocity profiles and the mean $I_{sp}$ values are displayed in figure \ref{fig:Isp1836}. The profiles have been smoothed for better readability.
Additionally, given the pulsed nature of reconnection, the $I_{sp}$ evolution over time is given in figure \ref{fig:IspTime_mr1836}
for the same cross sections considered earlier.

All the profiles show 
one or more peaks coinciding with the main outflow stream, 
whose amplitude is seen to decrease over time once the bulk outflow is past. 
After this point, reconnection is considered to be ended, and a new
reconnection event might eventually be expected to take place.
The positive values observed in the profiles (according the 
frame of reference here adopted) represent the particular ion dynamics due to the interaction with the magnetic islands formed along the process.
The particles recirculation generated within a magnetic island causes the velocity to achieve both positive and negative directions. 
Later than $36\ \unit{\omega_{c,i}^{-1}}$ the process is considered finished.

From the results in figure \ref{fig:IspTime_mr1836} the best outcome is achieved through the cross section at $x = 4\ \unit{d_i}$, 
which describes the situation at the outer distances. In this region the specific impulse reaches the 
highest values, and the curve shows a much more regular profile.

Additionally, we observe the peaks to be shifted leftwards and increased in magnitude over the cross sections to indicate an acceleration 
over the process. Noticeable is also the double peak observed through the cross-section in
$x = 6\ \unit{d_i}$, which points out a likely second acceleration of the ions at $22\ \unit{\omega_{c,i}^{-1}}$. The origin of this secondary acceleration is related to the occurrence of a secondary reconnection 
in the outflow. 
Indeed, panel (f) in figure \ref{fig:VxionsRuns1836}
shows the signature of an additional occurring reconnection event at $x \sim 5\ \unit{d_i}$, which may lead to explain 
such particles acceleration.

\subsection{Energy Evolution in Eulerian Frame}

\begin{figure}
 \begin{center}
  \includegraphics[scale=0.6]{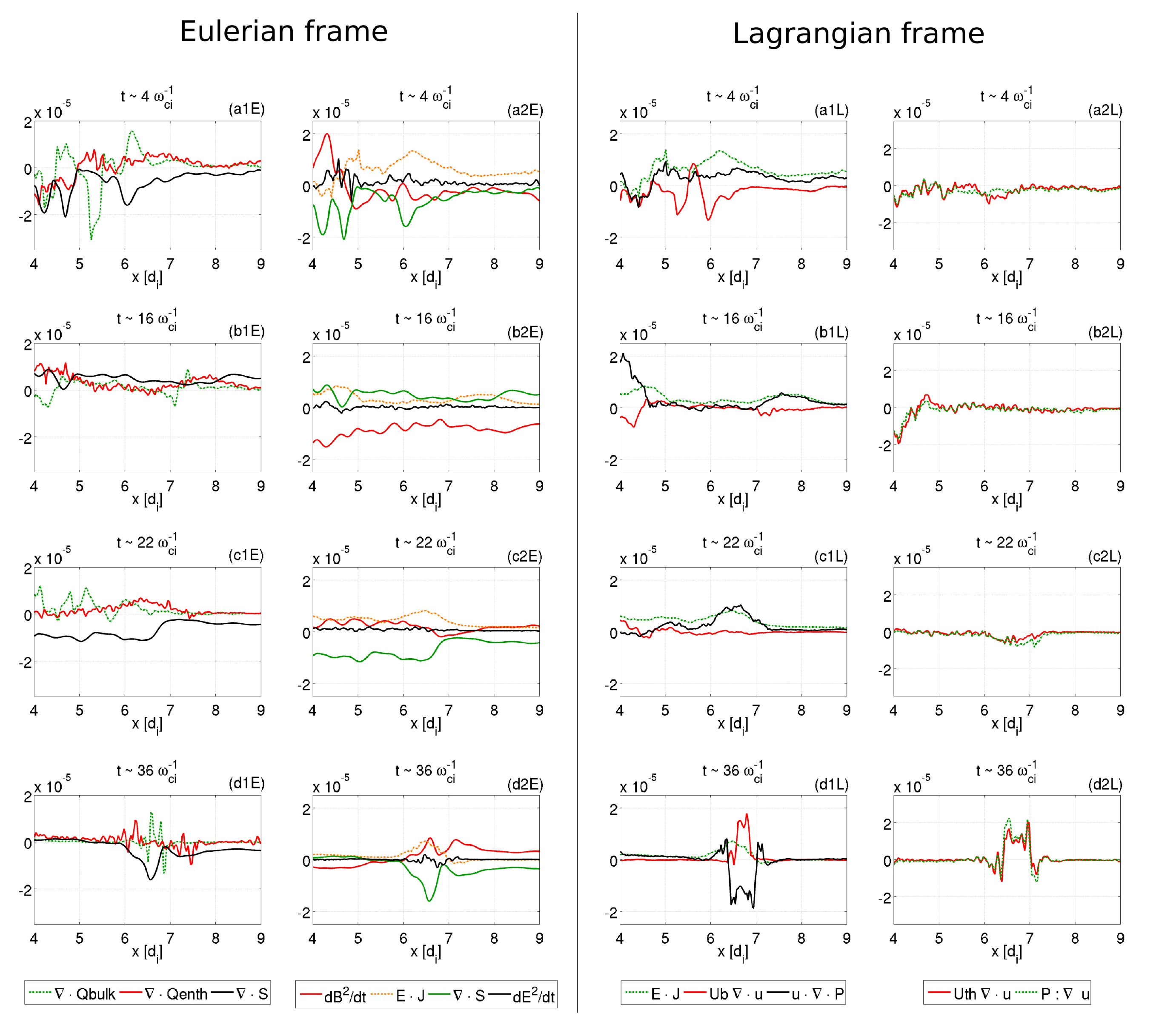}
    \caption{$y$-integrated profiles along the current sheet of the Eulerian (panels \emph{a1E - d1E}) and Lagrangian frame (panels \emph{a1L - d1L})
    at relevant time steps.}\label{fig:energyTOT}
 \end{center}
\end{figure}

This section aims at giving further insights into the global energy budget under this particular configuration.
So far, the budget in the case of symmetric reconnection has long been studied considering an Eulerian frame approach (e.g. \citet{birn2005, birn2009}), which reads 
\begin{equation} \label{eq:enTOT}
 \frac{\partial}{ \partial t } \left( \epsilon_0 E^2 + \frac{B^2}{\mu_0} + U_{th} + U_b \right) = - \nabla \cdot \left[ \mathbf{S} + \mathbf{Q}_{enth} + \mathbf{Q}_b + \mathbf{Q}_{th} \right]
  \end{equation}
  \begin{equation} \label{eq:electMagn}
 \frac{1}{2} \frac{\partial}{ \partial t } \left( \epsilon_0 E^2 + \frac{B^2}{\mu_0} \right) = - \mathbf{E} \cdot \mathbf{J} - \nabla \cdot \mathbf{S}
  \end{equation}
  where $\mathbf{S} = \frac{\mathbf{E} \times \mathbf{B}}{\mu_0}$ is the Poynting flux, while the other terms are the same as in equations \ref{eq:Qb}, \ref{eq:Qth} and \ref{eq:Q}, with additional term $\mathbf{Q}_{th}$
  due to the thermal energy flux.
  The latter is normally negligible and so will be considered in this work.
  Although being already considered in total energy equation \ref{eq:enTOT}, the electromagnetic balance is explicitly shown in  equation \ref{eq:electMagn} for a better analysis.
   The equation terms integrated over $y$ for the Eulerian case are shown in panels (a1E) - (d1E) of figure \ref{fig:energyTOT}.    

Four different time steps describing a difference situation are taken into account, by paying particular attention to the energy exchanged 
within the discontinuity structures revealed in figure \ref{fig:outflowAnalysis}.
Profiles have been smoothed for a better readibility.
The situation at $t \sim 36\ \unit{\omega_{c,i}^{-1}}$ (Fig. \ref{fig:energyTOT} panels \emph{d}), which displays a completely formed magnetic island, is taken as reference for the outflow-island interaction.
At $t \sim 4\ \unit{\omega_{c,i}^{-1}}$ a decrease in $\nabla \cdot \mathbf{S}$ corresponds to an increase in the bulk  and thermal
energy within the structures observed between $x = 6 \text{ and } 7\ \unit{d_i}$ in figure \ref{fig:outflowAnalysis}, indicating a particles acceleration and plasma heating 
at the expense of electromagnetic energy. 
Also, remarkable is the sudden $Q_b$ and $Q_{enth}$ reverse occurring downstream, i.e. between $4 \text{ and } 5\ \unit{d_i}$, caused by the outflow-island interaction, 
as further confirmed by panel (d1E).

Later in time, in correspondance of the \emph{diamond-chain} structure  
within the closed magnetic island at $t \sim 16\ \unit{\omega_{c,i}^{-1}}$, we observe
a situation similar to the previous step, with a slight  
$\nabla \cdot \mathbf{S}$ decrease followed by an increase in enthalpy, mainly limited to the earliest part of the profile. 
However, the bulk energy flux now shows a rapidly reversing profile, as the one observed in panel (d1E). The latter suggests 
the bulk energy is more influenced by the outflow-island interaction than
the discontinuity structure, which in turn is mainly acting as plasma heater.
Moreover, at the end of the outflow, at around $x = 4\ \unit{d_i}$, the presence of a dipolarization front is highlighted by an increase in the enthalpy and $\nabla \cdot \mathbf{S}$.
Finally, the situation at $t \sim 22\ \unit{\omega_{c,i}^{-1}}$ shows a completely different situation. The remarked electromagnetic energy  drop (i.e. $\nabla \cdot \mathbf{S}$) 
around $x \sim 7\ \unit{d_i}$ 
is now only followed by an increase in enthalpy, while $Q_b$ is seen to begin an oscillating escalation only later in the outflow. This latter behavior
coincides with the turbulent outflow region seen in figure \ref{fig:VxionsRuns1836}, where we observe an increment of the plasma thermal energy and a non-linear 
increase in the plasma bulk energy.

\subsection{Energy Evolution in Lagrangian Frame}

\begin{figure}
 \begin{center}
  \includegraphics[scale=0.35]{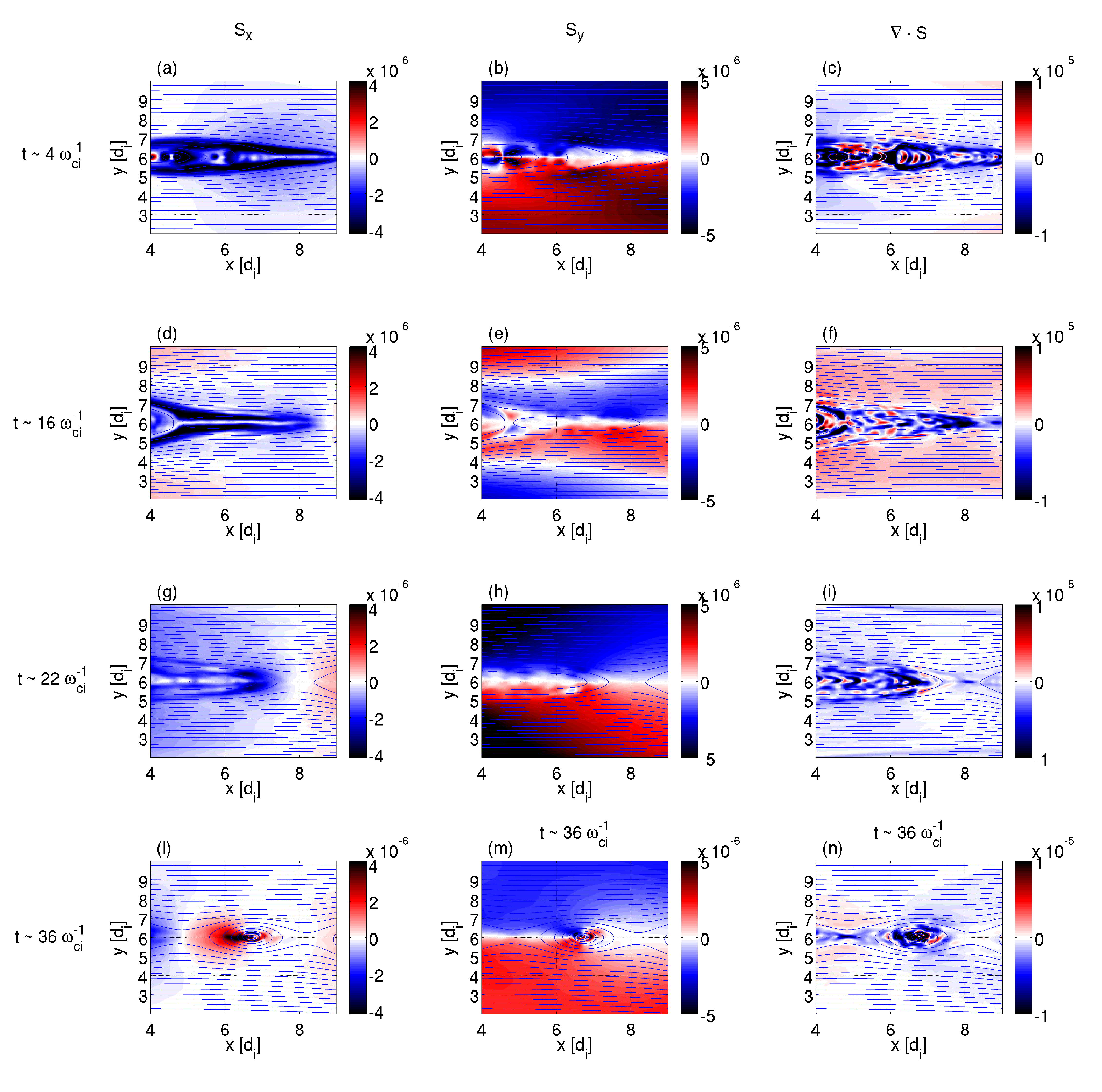}
    \caption{Plot of the $x$ and $y$ components of the Poynting vector, as well as its divergence $\nabla \cdot \mathbf{S}$, at the time steps considered for the energy analysis.
    The colorbar in code units.}\label{fig:poyn}
 \end{center}
\end{figure}

To gain more insights into the effective particle behavior within a moving fluid element, it is more interesting to analyse the energy budget by adopting a Lagrangian frame.
  The bulk and thermal energy equations are now threated separately, and read

  \begin{equation} \label{eq:bulkEnergy}
 \frac{D U_{b}}{D t} = \mathbf{E} \cdot \mathbf{J} - U_{b} \nabla \cdot \mathbf{u} - \mathbf{u} \cdot \nabla \cdot \mathbb{P}
\end{equation}

  \begin{equation} \label{eq:thermalEnergy}
 \frac{D U_{th}}{D t} = - U_{th} \nabla \cdot \mathbf{u} - \mathbb{P} : \nabla \mathbf{u}
\end{equation}
where the terms are the same as in the previous equations, in addition with  the heat flux $\mathbf{Q}_{h,f}$ (normally negligible), and the 
operator $\frac{D}{D t}$, which 
represents the total derivative. The $y$-integrated profiles are shown in panels (a1L) - (d2L) of figure \ref{fig:energyTOT} for the same time steps as the case 
in Eulerian frame.
In correspondance of the discontinuity structures at $t \sim 4\ \unit{\omega_{c,i}^{-1}}$, we now observe a dominance of $\mathbf{J} \cdot \mathbf{E}$ 
against a decrease in $\nabla \cdot \mathbf{S}$ and
the velocity divergence ($U_b \cdot \nabla \cdot \mathbf{u}$) to rapidly drop in favour of an increment 
of $\mathbf{E} \cdot \mathbf{J}$ and the pressure
tensor divergence ($u \cdot \nabla \cdot \mathbf{P}$).
Finally, from the thermal energy equation we understand that the discontinuity structure causes a slightly increase in  all the terms.
Interesting is the situation downtream the outflow at $t \sim 4 \unit{\omega_{c,i}^{-1}}$ and $t \sim 16\ \unit{\omega_{c,i}^{-1}}$, 
such as an inversion of $\mathbf{E} \cdot \mathbf{J}$ and a strong heating.  The latter is describing
the interaction of the outflow with an forming island.
Noticeable is also the situation described by equation \ref{eq:thermalEnergy}. All the terms are seen to follow the same pattern, with a large variation 
along the current sheet, except for
the case at  $t \sim 36\ \unit{\omega_{c,i}^{-1}}$. The latter shows an overall weak heating in the island edges, followed by a strong cooling in the inner region, 
likely due to an expanding evolution.
Finally, significant is the sharp increment of the thermal component observed at $t \sim 22\ \unit{\omega_{c,i}^{-1}}$ within the turbulent outflow region.

fFor a better understanding of the energy exchanged in the electromagnetic fields, 
we consider the evoluton of the Poynting vector, whose $x$ and $y$ components and its divergence are shown 
in Figure \ref{fig:poyn}. The component along $z$ has been left out as not contributing to the energy term $\nabla \cdot \mathbf{S}$.
On the whole, we notice that the $x$ component remains steadily negative over all the process. In symmetric reconnection
this component is mainly driven by the out-of-plane Hall magnetic field $B_z$ \citep{lapenta2013}, as it also happens in this case. 
However, notice that this situation reverses around $x \sim 4\ \unit{d_i}$ at $t \sim 4\ \unit{\omega_{c,i}}$, 
where a second reconnection event is taking place by generating a strong positive electromagnetic energy flux. The latter is governed by $E_x \text{ and } B_y$, 
with $E_z$ being dominating quantity in this region.
At the latest stage, a magnetic island is completely formed. The Poynting flux now signals a double $S_x$ polarity, where the positive part is clearly given by
a reconnection outflow shaping the island itself.

The component along $y$ also gives a good indication of reconnection. We observe a strong positive-negative energy flux 
all over outside the reconneciton region, 
and a void value in the outflow, until the encounter with the magnetic island. The energy flux is only dominated by the convective electric field 
$\mathbf{v} \times \mathbf{B}$. Being the inflow velocity
nearly along $x$ and the magnetic field principally along $x$, the Poynting component correctly indicates $E_z$ is the dominant term. 
No energy exchange is seen in the outflow, as expected given the peculiar  properties of the separatrices in shaping reconnection 
\citep{lapenta2014, lapenta2016}. Interesting is, however, the quadrupolar \emph{flower} structure observed in the island between $4 < x < 6\ \unit{d_i}$, which
is later destroyed as the process goes on. This pattern is  dominated by the $E_z \cdot B_x$ term, which in turn is seen to be governed by the 
field $B_x$ (plots not shown here).

Finally, $\nabla \cdot \mathbf{S}$ gives an insight into the energy flow magnitude. From Figure \ref{fig:poyn} we observe that this quantity is predominantly
negative, except for some reversed structures seen after the discontinuities analyzed earlier, where $\nabla \cdot \mathbf{S}$ becomes strongly positive.
The whole evolution is then predominantly 
driven by the gradient of $S_x$, except for the interaction with a forming island, whereby the $S_y$ component plays a more important role.
As the reconnection continues, the magnitude  becomes particularly irregular, followed by a more collimated pattern along the outflow direction, 
probably due to secondary reconnection events taking over (e.g. at $t \sim 22\ \unit{\omega_{c,i}}$).

\begin{figure}
 \begin{center}
  \includegraphics[scale=0.50]{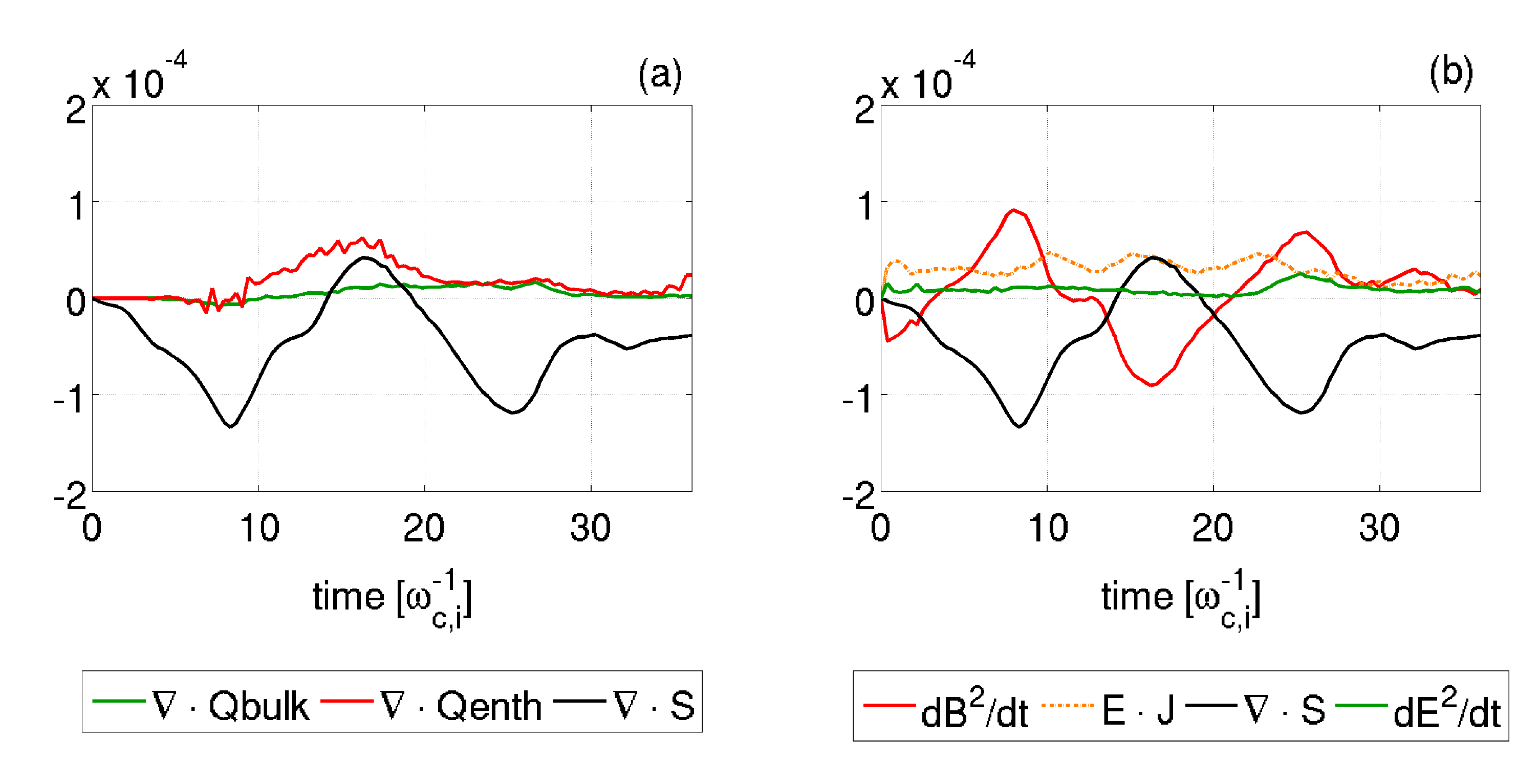}
    \caption{Temporal profiles of the energy equation terms in the Eulerian frame (Eqs. \ref{eq:enTOT} and \ref{eq:electMagn}). All terms are introduced in the text.}\label{fig:energyDom}
 \end{center}
\end{figure}

As final analysis, we propose in Figure \ref{fig:energyDom} the temporal evolution of the terms in the energy equations 
\ref{eq:enTOT} and \ref{eq:electMagn}, as integrated over the all domain. To represent the single outflow energetics 
the values are halved thanks to the symmetry of the system with respect to the X-point. We focus only on the 
Eulerian approach to better understand the energy flows across the whole system.
On the whole, we notice the electromagnetic profiles, including $\nicefrac{\partial  B^2}{\partial t}$ and $\nabla \cdot \mathbf{S}$, 
to show a certain degree of periodicity with period of nearly $\omega_{c,i} \cdot t \sim 17$.
 This fact can be interpreted as reconnection to be still at its unsteady stage, with the steadiness attained after $ 30\ \unit{\omega_{c,i}^{-1}}$. 
 Accordingly, after this point all the temporal terms 
 on the lefthand side of the equation are nulled.
From panel (a),
we notice a certain intial delay 
before the initial electromagnetic energy is converted into bulk and enthalpy energy (nearly $ \sim 8\ \unit{\omega_{c,i}^{-1}}$). After this initial latency, the total energy is mainly 
converted to both bulk and enthalpy energy, with the latter receiving the greater share. The enthalpy is seen to rapidly grow in the range 
$10 - 17\ \unit{\omega_{c,i}^{-1}}$ and sharply drop afterwards, meanwhile the bulk energy continues to slightly grow with a lower rate until the steady state is reached.
Plasma is therefore firstly rapidly heated and then mildly constantly  accelerated.
From panel (b) we observe that the most dominant terms over the unsteady period are the temporal variation of the magnetic energy and the Poynting vector flux, both following a pulsed behavion.
Moreover, we notice as the latter obtains a positive value between $ 15 \text{ and } 18\ \unit{\omega_{c,i}^{-1}}$, to indicate that some electromagnetic energy has been earlier accumulated and later
released to build the EM up again. This EM build-up results typical of reconnection and positively concur to keep the partcle energy state significantly high, as confirmed by the jumps in the $\mathbf{E} \cdot \mathbf{J}$ 
profile.

\section{\label{sec:conclusions} Conclusions}

This work aimed at analysing the ion dynamics from fully kinetic simulations of symmetric magnetic reconnection in typical laboratory plasma configurations.
Simulations were performed using the massively parallel fully-kinetic implicit-moment Particle-in-Cell code iPIC3D \citep{markidis2010}.
A not-force free unbalanced profile is adopted across the current sheet to significantly accelerate the reconnection process and at the same time allow for a free choice of the initial setup.
A set of nine simulations with reduced mass ratio $m_r = 512$ has been performed by permuting three significant values of the magnetic field ($B = 200\ \unit{G}$, $800\ \unit{G}$, $5000\ \unit{G}$),
and electron temperature ($T_e = 0.5$, $3$, $10\ \unit{eV}$).
The initial ion temperature and density are kept constant at $T_i = 0.0215\ \unit{eV}$ (i.e. room temperature) and $10^{19}\ \unit{m^{-3}}$, respectively.
The flow velocity of the ions outside the reconnection region has been analyzed, showing that the reconnection process can efficiently accelerate ions up to velocities comparable with the inflow
Alfv\'en speed $V_A$ even in laboratory conditions (nearly $0.60$-$0.70\ \unit{V_A}$). 
Finally, the configuration resulting in the most promising acceleration has been simulated with a realistic mass ratio.
From the analysis of the enegy budget shared during reconnection, we observe that nearly $30\ \unit{\%}$ of the initial energy is given to particles, from which great part of it is subsequently 
converted into heat. After $t \sim 25\ \unit{\omega_{c,i}}$ we observe a slightly increase in the magnetic field energy, which is thought to be caused by the formation of a secondary island. 
In correspondance of this increase, the total particles energy shows a mild drop, with most of this energy being thermal energy.
Interestingly, this simulation revealed the formation of a series of discontinuities over the reconnection outflow, as shown in figure \ref{fig:outflowAnalysis}. 
Similar structures were observed in 
\textcite{zenitani2011} and \textcite{zenitani2015} and in relativistic simulations, which are believed to be shock 
structures. Additional remarkable features are also observed, such as the series of complex structures visible between $x = 5$ and $x = 7\ \unit{d_i}$ 
in the left panel.
Although further investigation is required to fully understand their nature, we can already observe and quantify those 
dissipative mechanisms causing an 
increase in both temperature and entropy at the expenses of ions directional (kinetic) energy.
The latter is further confirmed by the Eulerian and Lagrangian energy analysis, which show these two type of structures to increase the bulk and enthalpy energy flux at the expense of electromagnetic energy.
Similarly, the thermal energy equation in Lagrangian frame mostly shows that an increase in enthalpy and velocity divergence is observed within these regions.

Finally, an important goal of this work was to figure out the potentiality of magnetic reconnection in accelerating particles for industrial and space propulsion purposes.
Such outcomes in the ion outflow velocity suggests possible applications in the domain of spacecraft propulsion.
As such, a parameter similar to the one used in electric propulsion, i.e. the specific impulse $I_{sp}$, was taken as quality parameter.
Its average value across different simulation box cross sections is evaluated showing very interesting 
results. 
The maximum $I_{sp}$ with a hydrogen plasma is as high as $330 \cdot 10^{3}\ \unit{s}$, attained at locations relatively far from the reconnection region and after reconnection is fully developed.

In conclusion, we argue that 
magnetic reconnection is surely worth being further studied in light of specific applications for 
particles acceleration. Given its remarkable specific impulse, as well as the high velocity values
reached in the outflow, this process appears to be suitable to applications like spacecraft propulsion and generation of ion beams.

\begin{acknowledgments}
 
The present work has been possible thanks to the 
Illinois-KULeuven Faculty/PhD Candidate Exchange Program.
This research has received funding from the Onderzoekfonds KU Leuven (Research Fund KU Leuven) and 
by the Interuniversity Attraction Poles Programme of the Belgian Science Policy Office (IAP P7/08 CHARM). 
The simulations were conducted on the computational resources provided by the PRACE Tier-0 machines (Curie, Fermi and MareNostrum III supercomputers), under the PRACE Project 2010PA2844. 
 
\end{acknowledgments}
 
%

\end{document}